\pdfoutput=1

\documentclass[11pt]{article}

\usepackage[final]{acl}

\usepackage{times}
\usepackage{latexsym}

\usepackage[T1]{fontenc}

\usepackage[utf8]{inputenc}

\usepackage{microtype}

\usepackage{inconsolata}

\usepackage{graphicx}

\usepackage{xcolor}

\title{A General Pseudonymization Framework for Cloud-Based LLMs: \\Replacing Privacy Information in Controlled Text Generation}

\author{
 \textbf{Shilong Hou\textsuperscript{1}},
 \textbf{Ruilin Shang\textsuperscript{1}},
 \textbf{Zi Long\textsuperscript{2,}\thanks{Corresponding author}},
 \textbf{Xianghua Fu\textsuperscript{2}},
 \textbf{Yin Chen\textsuperscript{2}}
\\
\\
 \textsuperscript{1} College of Application and Technology, Shenzhen University, China
 \\
 \textsuperscript{2} College of Big Data and Internet, Shenzhen Technology University, China
\\
 \small{
   \textbf{Correspondence:} \href{mailto:longzi@sztu.edu.cn}{longzi@sztu.edu.cn}
 }
}

\begin{document}
\maketitle

\begin{abstract}
\label{sec:abs}
An increasing number of companies have begun providing services that leverage cloud-based large language models (LLMs), such as ChatGPT. 
However, this development raises substantial privacy concerns, as users' prompts are transmitted to and processed by the model providers.
Among the various privacy protection methods for LLMs, those implemented during the pre-training and fine-tuning phrases fail to mitigate the privacy risks associated with the remote use of cloud-based LLMs by users.
On the other hand, methods applied during the inference phrase are primarily effective in scenarios where the LLM's inference does not rely on privacy-sensitive information.
In this paper, we outline the process of remote user interaction with LLMs and, for the first time, propose a detailed definition of a general pseudonymization framework applicable to cloud-based LLMs.
Building upon the framework, we have designed various pseudonymization methods and further propose a method that achieves pseudonymization through a controllable text generation process.
The experimental results demonstrate that the proposed framework strikes an optimal balance between privacy protection and utility.
The code for our method is available to the public at \url{https://github.com/Mebymeby/Pseudonymization-Framework}.
\end{abstract}

\section{Introduction}
\label{sec:intro}

\begin{figure}[t]
    \centering
    \includegraphics[scale=0.28]{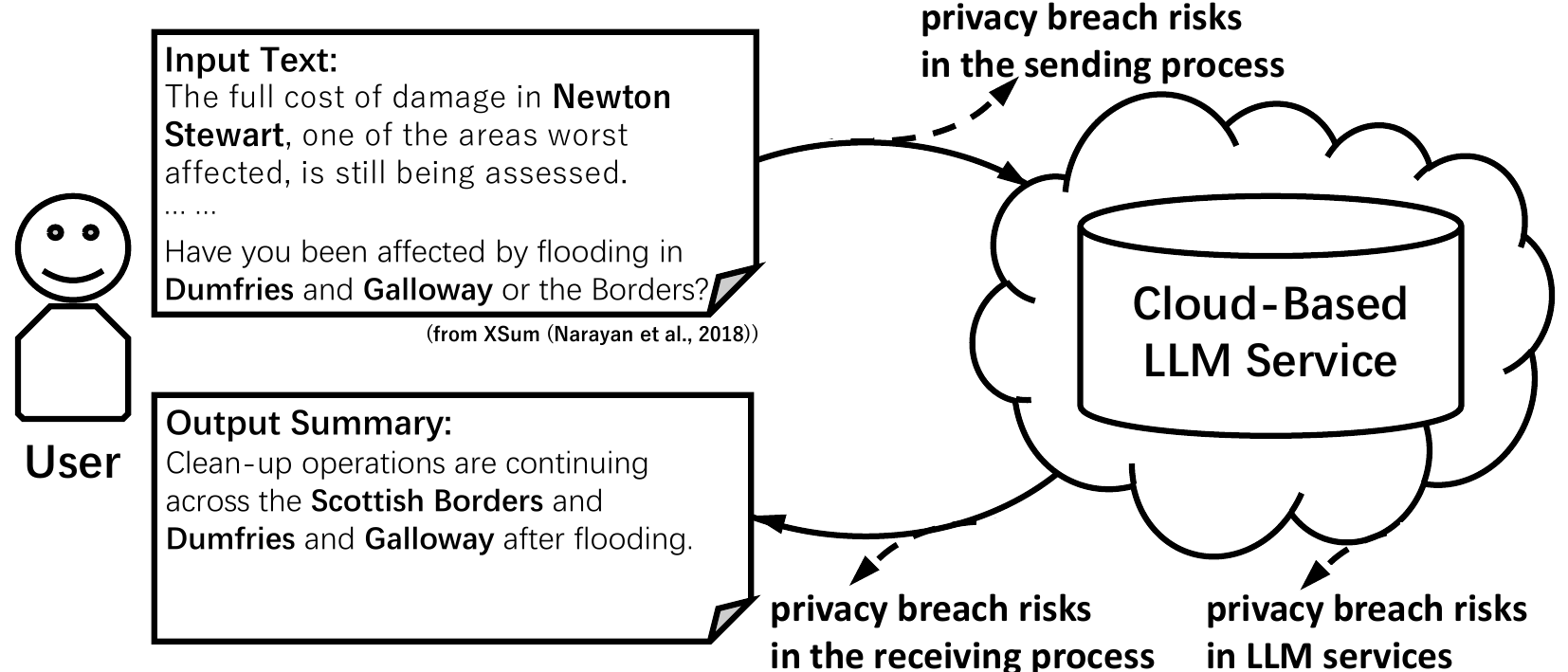}
    \caption{Potential privacy breach risks in using cloud-based LLM services}
    \label{fig:problem}
\end{figure}

Large Language Models (LLMs) have demonstrated considerable promise in advancing the field of artificial intelligence, showcasing remarkable capabilities in instruction following and excelling across a wide range of tasks, including writing, coding, and other text-based activities~\citep{bubeck2023sparks,touvron2023llama,openai2024gpt4technicalreport}.
Consequently, an increasing number of companies have begun providing cloud-based LLM services, such as ChatGPT\footnote{\url{https://chatgpt.com/}}.
However, the widespread use of cloud-based LLM services has raised substantial privacy concerns: the transmission and storage of user data on cloud infrastructures pose significant risks of data breaches and unauthorized access to private information, as illustrated in Figure~\ref{fig:problem}.

Current privacy-preserving techniques for cloud-deployed LLMs either prevent untrustworthy customers from accessing privacy-sensitive information in pre-trained datasets~\citep{carlini2019secret,pan2020privacy,brown2022does}, or safeguard users' pre-training and fine-tuning datasets from untrustworthy cloud service providers~\citep{chi2018privacy,jegorova2022survey}.
However, these methods face significant challenges in addressing the unique issues arising from remote access to cloud-based LLMs. 
On the other hand, researchers have developed various strategies to ensure privacy security during the inference phrase, including Multi-Party Computation~\citep{goldreich1998secure}, homomorphic encryption~\citep{acar2018survey}, differential privacy in inference~\citep{majmudar2022differentially}.
However, these methods are not suitable for scenarios in which the cloud-based LLM's inference relies on privacy-sensitive information.

The data pseudonymization technique, which ensures privacy protection by appropriately replacing privacy-sensitive information, has since attracted the attention of researchers.~\citep{kan2023protecting,chen2023hide,lin2024emojicrypt}
However, research on applying pseudonymization techniques during the inference phase for privacy protection remains limited. 
Currently, a detailed definition of a pseudonymization framework for the inference phase of cloud-based LLMs is lacking.
For example, \citet{yermilov2023privacy} divides pseudonymization into two parts: recognizing and replacing privacy entities. 
However, \citet{chen2023hide} argues that pseudonymization should consist of two stages: concealing privacy entities for anonymization and restoring them for de-anonymization.
We argue that these methods integrate certain steps of the pseudonymization process and, therefore, cannot be regarded as a general pseudonymization framework.



In this paper, we outline the process of remote user interaction with LLMs and, for the first time, propose a detailed definition of a general pseudonymization framework applicable to cloud-based LLMs.
We define the pseudonymization framework as comprising three components: the detection of privacy-sensitive information, the generation of replacement terms, and the replacement of privacy information to achieve pseudonymization.
We further propose a pseudonymization method based on a controllable text generation process, ensuring that the replaced text preserves maximal semantic correctness after replacement.
Furthermore, to evaluate the practical effectiveness of the proposed framework in real-world LLM services, we specifically assessed its performance in text generation tasks, including summarization, question answering, text generation, and machine translation, in addition to classification tasks. 
The experimental results indicate that the proposed framework achieves an optimal balance between privacy protection and utility.

To summarize, our contributions are as follows:
\begin{itemize}
    \setlength{\itemsep}{0pt}
    \item[(1)] We propose a general pseudonymization framework applicable to cloud-based LLMs.
    \item[(2)] We propose a pseudonymization method leveraging a controllable text generation process to preserve the semantic integrity of the replaced text.
    \item[(3)] We evaluate the proposed framework across various text generation tasks and demonstrate that it achieves the optimal balance between privacy and performance.
\end{itemize}

\section{Related Works}
\label{sec:rel}

Privacy protection for large language models (LLMs) can be categorized according to the phase in which it is implemented: during the pre-training and fine-tuning phases, and during the inference phase~\citep{yan2024protecting}.
Privacy protection during the pre-training and fine-tuning phases of LLMs is essential for safeguarding sensitive data while preserving model effectiveness. 
Techniques such as differential privacy~\citep{li2021large,wu2022adaptive,xu2024fwdllm}, data cleaning~\citep{bai2022training,kandpal2022deduplicating}, and federated learning~\citep{yu2023federated,xu2024fwdllm,zhang2024towards} can be utilized to mitigate privacy risks during these phases.
As previously discussed, these methods primarily aim to protect the privacy of information within LLMs. However, they do not fully address the privacy concerns associated with remote access to LLM services. 
Additionally, privacy protection measures implemented by model providers may not completely alleviate users' concerns regarding the potential misuse of their private data by these providers.

On the other hand, the issue of privacy leakage during the inference phase of LLMs has garnered significant attention.
To address this issue, researchers have developed numerous strategies to ensure privacy security during the inference phase. 
These include encryption-based privacy protection approaches such as Multi-Party Computation~\citep{goldreich1998secure, dong2022fusion}, homomorphic encryption~\citep{acar2018survey, hao2022iron,lu2023bumblebee}, and differential privacy in inference~\citep{dwork2006differential, dwork2008differential, majmudar2022differentially}.
%
For example, \citet{huang2022cheetah} proposed a specialized encoding method, Cheetah, which encodes vectors and matrices into homomorphic encryption polynomials.
However, these homomorphic encryption methods are challenging to apply to cloud-based black-box LLMs, as they require access to the model's internal structures.
%
Additionally, \citet{du2023dp} introduced DP-Forward, which applies differential privacy during inference by perturbing embedding matrices in the forward pass of language models. 
However, these differential privacy approaches are mainly effective when the LLM’s decision-making does not rely on sensitive information, which differs from the focus of our research.

\begin{figure*}[th]
    \centering
    \vspace{-0.2cm}
    \includegraphics[scale=0.37]{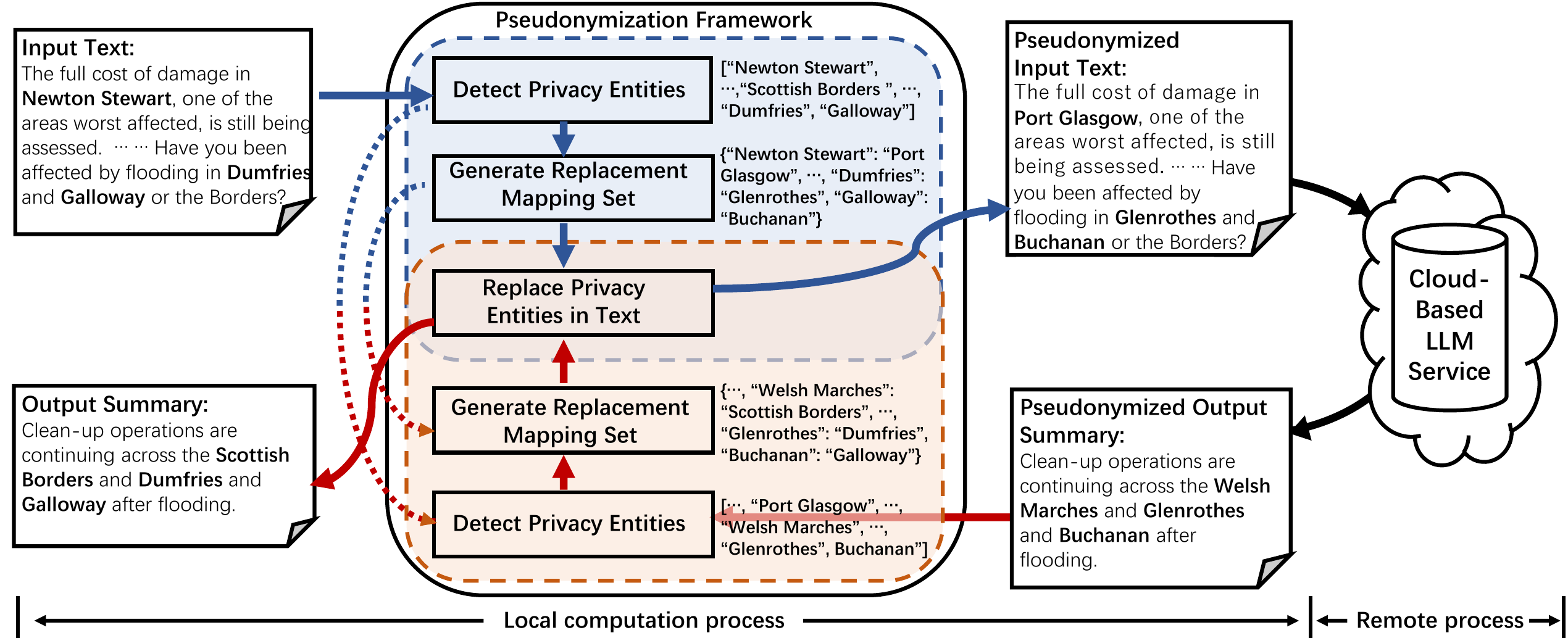}
    \caption{Overview of pseudonymization framework for cloud-based LLMs}
    \vspace{-0.2cm}
    \label{fig:framework}
\end{figure*}

In addition to the aforementioned methods, pseudonymization techniques focus on safeguarding the privacy of the prompt by identifying and removing privacy-sensitive information.
For example, \citet{kan2023protecting} and \citet{chen2023hide} proposed anonymizing sensitive terms before inputting them into the LLM and restoring them after the output.
\citet{lin2024emojicrypt} proposed a pseudonymization method to safeguard user privacy by converting user input from natural language into a sequence of emojis.
\citet{zhang2024cogenesis} introduced a mixed-scale model collaboration approach that combines the strengths of a large cloud-based model with a smaller, locally deployed model.
However, there is currently no general definition of a pseudonymization framework for the inference phase of cloud-based LLMs.
Additionally, these methods have primarily been tested on classification tasks, which differ from the core task of text generation in LLMs. Therefore, their results may not fully capture their effectiveness in text generation.


\section{Pesudonymization Framework}
\label{sec:framework}

As shown in Figure~\ref{fig:framework}, a privacy-preserving cloud-based LLM access process consists of two steps: pseudonymizing the privacy information in the input text, as indicated by the blue arrow, and restoring the privacy information in the output results, as indicated by the red arrow.
It is clear that the pseudonymization and restoration processes are logically identical, involving the detection of information to be replaced (e.g., privacy entities or entities to be restored), the generation of replacement candidates for detected entities, and the execution of the replacement process.
Furthermore, the detection and candidate generation in the restoration process can refer to the results of the pseudonymization process, while the replacement operation itself is identical to that in the pseudonymization process.
Therefore, we propose that a general pseudonymization framework should include only the three components of detection, generation and replacement.
In the following sections, we will provide a detailed definition of the tasks for each component and discuss several viable approaches for each stage.

\subsection{Detecting Privacy Information}
\label{sec:detect}
Given a user's input \( X \), which may contain multiple pieces of private information, we denote these pieces as \( P = \{ p^j_{A_i} | p^j_{A_i} \in X, 1 \leq i \leq n, 1 \leq j \leq N_i \} \). 
Here, \( A_i \) represents the \( i \)-th privacy attribute (e.g., name, location), and each \( p^j_{A_i} \) represents the \( j \)-th instance of private information related to the attribute \( A_i \). 
The total number of private information entries related to \( A_i \) is denoted as \( N_i \). 
The goal of the privacy information detection method is to collect \( P' = \{ p'^j_{A_i} | p'^j_{A_i} \in X, 1 \leq i \leq n, 1 \leq j \leq N_i \} \), where \( P' \) represents the collection of detected private information. 
To maximize security, \( P' \) should closely approximate \( P \), ensuring that all relevant private information is correctly identified while minimizing the risk of missing any sensitive data.
The three detection methods employed in our experiments are described as follows.

    \textbf{NER-based Detection} uses an off-the-shelf NER system to identify spans of named entities that correspond to privacy information categories. In this work, we utilize the publicly available BERT model, bert-large-cased-finetuned-conll03-english~\footnote{\url{https://huggingface.co/dbmdz/bert-large-cased-finetuned-conll03-english}}.
    We refer to this method as \( \mathrm{DET_{NER}} \).
    
    \textbf{Prompt-based Detection} employs a locally deployed, small-scale instruction-tuned LLM to identify named entities. 
    We denote this method as \( \mathrm{DET_{prompt}} \).
    
    \textbf{Seq2Seq Detection} is developed by fine-tuning a small-scale base LLM on a parallel corpus of pseudonymized texts generated using the NER-based detection method. This method generates sentences that maintain consistency with the input text while marking or replacing privacy entities with designated tags, as illustrated in Table~\ref{tab:exp_seq2seq_det}. We denote the two Seq2Seq detection variants as \( \mathrm{DET_{tag\_mark}} \) and \( \mathrm{DET_{tag\_rep}} \).

\begin{table}[t]
    \centering
    \begin{tabular}{ll}
        \hline
        \!\!\!\textbf{Input}\!\!\! & John Edward Bates, formerly of \\
        & Spalding, is now living in London. \\ \hline
        \!\!\!\textbf{Output}\!\!\! & \texttt{<ENT>}John Edward Bates\texttt{</ENT>}, \\ 
         \!\!\!\textbf{(mark) }\!\!\!  & formerly of \texttt{<ENT>}Spalding\texttt{</ENT>}, \\ 
         &\!\!\!is now living in \texttt{<ENT>}London\texttt{</ENT>}.\!\!\!\\ \hline
        \!\!\!\textbf{Output}\!\!\! & \texttt{<ENT>}, formerly of \texttt{<ENT>}, is now \\
         \!\!\!\textbf{(replace)}\!\!\! & living in \texttt{<ENT>}.  \\
        \hline
    \end{tabular}
    \caption{Example output of Seq2Seq detection with entity marking and replacement}
    \vspace{-0.2cm}
    \label{tab:exp_seq2seq_det}
\end{table}

\subsection{Generating Replacement Candidates}
\label{sec:gen}

Based on the detected privacy entities \( P' \), the next step is to generate candidate entities \( Q \) that do not contain any privacy information to replace \( P' \). 
Specifically, the goal of generation is to obtain a replacement mapping set \( \mathcal{P} = \{ (p'^j_{A_i}, q^j_{A_i}) | p'^j_{A_i} \in X, 1 \leq i \leq n, 1 \leq j \leq N_i \} \), where \( q^j_{A_i} \) represents the generated candidate for \( p'^j_{A_i} \).
To ensure that the meaning of the original sentence remains intact after replacement, the replaced entities should generally share certain common characteristics (e.g., gender and language for names) with the original entities.
Building on the aforementioned requirement, the semantics of \( p'^j_{A_i} \) and \( q^j_{A_i} \) should be as distinct as possible, ensuring that privacy information cannot be easily inferred from \( q^j_{A_i} \).
%
The two candidate generation methods employed in our experiments are described as follows.

    \textbf{Random Sampling} utilizes the entities identified in Section~\ref{sec:detect} as a candidate set. From this set, an entity belonging to the same category as the privacy entities to be replaced is randomly selected as the replacement candidate. We denote this method as \( \mathrm{GEN_{rand}} \).
    
    \textbf{Prompt-based Generation} employs a locally deployed, small-scale instruction-tuned LLM to generate replacement candidates for the privacy entities. 
    We denote this method as \( \mathrm{GEN_{Prompt}} \).


\subsection{Replace Privacy Entities}
\label{sec:replace}
Given the input text \( X \) and the replacement mapping set \( \mathcal{P} \) obtained from the previous sections, the next step is to replace the entity \( p'^j_{A_i} \) in \( X \) with the corresponding replacement entity \( q^j_{A_i} \). 
The resulting text after replacement is denoted as \( X' \).
To ensure that the meaning of the original text is preserved after the replacement, the remaining content in the text, aside from the replaced entities, should be appropriately adjusted. 
In other words, the goal of privacy entity replacement is to ensure that \( X' \) retains as much semantic correctness as possible. 
\( X' \) is then processed through a prompt template function and input into cloud-based LLMs, generating the output \( Y' \).  
As mentioned earlier, for \( Y' \), there is no need to perform privacy entity detection and replacement candidate generation. Instead, the restoration process of \( Y' \) involves directly replacing \( q^j_{A_i} \) in \( Y' \) with \( p'^j_{A_i} \), similar to the replacement process in \( X \), resulting in the final output \( Y \).
The three entity replacement methods employed in our experiments are described as follows.

    \textbf{Direct Replacement} refers to the process of directly replacing \( p'^j_{A_i} \) with \( q^j_{A_i} \) without modifying other parts of the text \( X \). This method is denoted as \( \mathrm{REP_{direct}} \). As previously mentioned, this approach may introduce semantic errors.
    
    \textbf{Prompt-based Replacement} employs a locally deployed, small-scale instruction-tuned LLM to perform the replacement of entity names. 
    We denote this method as \( \mathrm{REP_{prompt}} \).
    
    \textbf{Replacement through Text Generation} executes replacement during a controllable text generation process to ensure the semantic correctness of the text after replacement.
    When the detected privacy entity term \( p'^j_{A_i} \) is encountered during the text generation process, it is replaced by the corresponding entity \( q^j_{A_i} \), and the generation of the subsequent token proceeds accordingly. 
    The specific technical details of this method will be discussed in Section~\ref{sec:pesu_method}. 
    We denote this method as \( \mathrm{REP_{gen}} \).

\section{Pesudonymization Through Controllable Text Generation}
\label{sec:pesu_method}

We propose a pseudonymization replacement method based on a controllable text generation process, ensuring that the replaced text preserves maximum semantic correctness.
In this section, we provide a detailed explanation of the method's process.

\begin{figure}[t]
    \centering
    \vspace{-0.1cm}
    \includegraphics[scale=0.38]{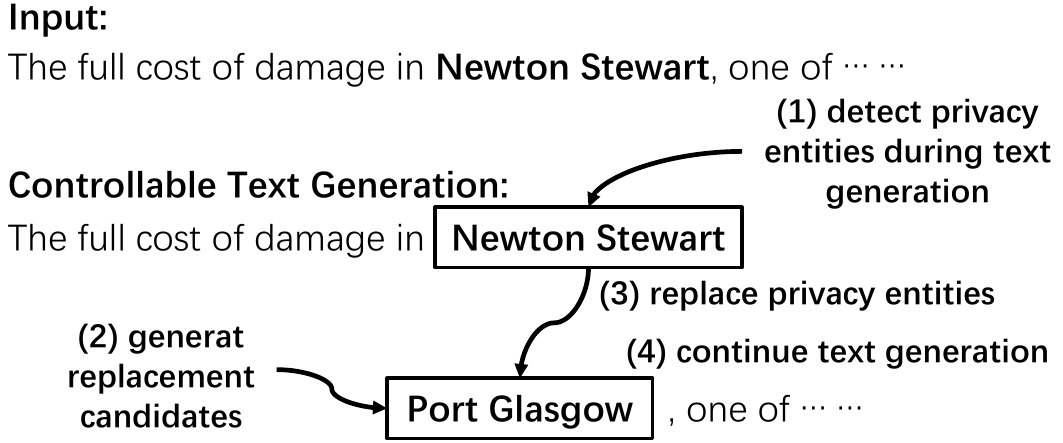}
    \caption{Workflow of pesudonymization through controllable text generation}
    \vspace{-0.2cm}
    \label{fig:generation}
\end{figure}

Given \( X = (x_1, x_2, \dots, x_L) \), the generation process of the LLM can be formulated as a sequential prediction of the next token, expressed as follows:  
\[
\hat{y}_{i} = \mathrm{argmax} \ P(y_{i} \mid g(X), \hat{y}_1, \dots, \hat{y}_{i-1})
\]  
Here, \( g(X) \) represents the prompt text generated from \( X \) using a predefined prompt template, and \( \hat{y}_{i} \) (where \( 1 \leq i \leq N \)) denotes the predicted token at the \( i \)-th time step.
As illustrated in Figure~\ref{fig:generation}, during the pseudonymization process, the majority of the output text \( \hat{Y} = (\hat{y}_1, \dots, \hat{y}_N) \) remains identical to the input text \( X \), except for a small portion where privacy entities is replaced.

Note that when using NER-based or prompt-based detection methods to identify privacy entities, we first employ a model capable of generating text identical to the input. During the text generation process, we compare each generated token \( \hat{y}_{i} \) with elements in \( P' \) to determine whether \( \hat{y}_{i} \) corresponds to a privacy entity.
Therefore, depending on the privacy entity detection method used, \( \hat{y}_{i} \) can take the following forms:  
\begin{description}  
\setlength{\itemsep}{0pt}  
    \item[(1)] \( \hat{y}_{i} = x_i \), where \( x_i \notin P' \). Here, \( P' \) represents a set of identified privacy entities collected using NER-based or prompt-based detection methods, as described in Section~\ref{sec:detect}.  
    \item[(2)] \( \hat{y}_{i} = x_i \), where \( x_i \in P' \). In this case, \( x_i \) is recognized as a privacy entity by the NER-based or prompt-based detection methods.  
    \item[(3)] \( \hat{y}_{i} = x_i \) when utilizing the Seq2Seq detection method described in Section~\ref{sec:detect}.  
    \item[(4)] \( \hat{y}_{i} = \texttt{<ENT>} x_i \texttt{</ENT>} \) or \( \hat{y}_{i} = \texttt{<ENT>} \). In this case, \( x_i \) is recognized as a privacy entity by the Seq2Seq detection method.  
\end{description}

Next, for privacy entity \( x_{i} \) in cases (2) or (4), we generate the replacement candidate \( x'_{i} \) corresponding to \( x_{i} \), based on the method described in Section~\ref{sec:gen}. 
Then, we set \( \hat{y}'_{i} = x'_{i} \).
As shown in Figure~\ref{fig:generation}, \( \hat{y}'_{i} \) will be incorporated into the above formula, and the prediction for the output at the \( (i+1) \)-th time step will proceed as follows:

\[
\hat{y}_{i+1} = \mathrm{argmax} \, P(y_{i+1} | \, g(X), \hat{y}_1, \dots, \hat{y}_{i-1}, \hat{y}_{i}')
\]
This process continues until the entire sequence has been generated.

The main contribution of this method lies in its ability to decouple the end-to-end pseudonymization text generation process\footnote{
In our preliminary experimental results, methods for pseudonymization through end-to-end text generation, such as those proposed by \citet{yermilov2023privacy} and \citet{chen2023hide}, yielded catastrophic results when trained with a limited amount of training data.
} into the three distinct stages described in Section~\ref{sec:framework}.
Additionally, it achieves better pseudonymization results by integrating different methods.
By performing pseudonymization through the controllable text generation process, this approach ensures comprehensive coverage of privacy information detection and the correctness of replacement candidate generation by integrating various detection and generation methods. 
Furthermore, this approach leverages the strengths of LLMs and Seq2Seq generation processes, maximizing the semantic correctness of the text after replacement.

\section{Experiment}
\label{sec:exp}

\subsection{Experiment Settings}
\label{sec:setting}



\begin{table*}[th]
    \centering
    \vspace{-0.2cm}
    \begin{tabular}{|l||c|c|c|c|c|c|}
    \hline
     methods & SQuAD & XSum & CNN/ & SAMSum & GLUE &  WMT14 \\ 
       & 2.0 & & Dailymail & & (MNLI) & (de-en) \\ \hline \hline
     Qwen2.5-14B  & F1 = 79.1 & ROUGE- & ROUGE- & ROUGE- & ACC =  &  BLEU =  \\ 
     -Instruct & EM = 75.5 & 1/2/L = & 1/2/L =  & 1/2/L =  & 84.3 &  12.2 \\ 
       &  &\!\!\! 25.4/7.0/17.8 \!\!\!& \!\!\! 30.8/10.2/20.5 \!\!\! & \!\!\! 41.9/15.8/32.8 \!\!\! &  &  \\ \hline \hline
     Qwen2.5-1.5B & F1 = 58.6 & ROUGE- & ROUGE- & ROUGE- & ACC =  & BLEU =  \\ 
     -Instruct  & EM = 55.4 & 1/2/L =  & 1/2/L = & 1/2/L = & 69.9 & 8.0 \\ 
      & & \!\!\! 18.9/3.8/13.2 \!\!\! & \!\!\! 23.7/7.8/16.5 \!\!\! &\!\!\! 36.4/13.0/28.5 \!\!\!& & \\ \hline \hline
     \( \mathrm{DET_{NER}} \)  & F1 = \textbf{76.6} & ROUGE- & ROUGE- & ROUGE- & ACC = & BLEU =  \\ 
     \( + \mathrm{GEN_{rand}} \)  & EM = \textbf{73.0} & 1/2/L =  & 1/2/L = & 1/2/L = & 81.6 & 9.9 \\ 
     \( + \mathrm{REP_{direct}} \)  & & \!\!\! 22.5/4.5/15.3 \!\!\! & \!\!\! 28.3/8.7/18.9 \!\!\! &\!\!\! \textbf{41.0/15.2/32.1} \!\!\!& &  \\ \hline
     \( \mathrm{DET_{NER}} \)  & F1 = 75.7 & ROUGE- & ROUGE- & ROUGE- & ACC = & BLEU =  \\ 
     \( + \mathrm{GEN_{prompt}} \)  & EM = 71.2 & 1/2/L =  & 1/2/L = & 1/2/L = & \textbf{83.0} & 9.5 \\ 
     \( + \mathrm{REP_{direct}} \)  & &\!\!\! \textbf{23.0}/4.9/15.8 \!\!\! &\!\!\! 28.8/8.7/19.2 \!\!\!&\!\!\! 40.7/\textbf{15.2}/31.9 \!\!\!& &  \\ \hline 
     \( \mathrm{DET_{prompt}} \)  & F1 = 74.8 & ROUGE- & ROUGE- & ROUGE- & ACC = & BLEU =  \\ 
     \( + \mathrm{GEN_{prompt}} \)  & EM = 70.9 & 1/2/L =  & 1/2/L = & 1/2/L = & 80.0 &  9.2 \\ 
     \( + \mathrm{REP_{prompt}} \)  & & \!\!\! 22.9/\textbf{5.7}/\textbf{15.9} \!\!\! & \!\!\! 24.4/7.1/16.3 \!\!\! & \!\!\! 32.3/11.3/25.5 \!\!\! & &  \\ \hline 
     \( \mathrm{DET_{NER}} \)  & F1 = 66.5 & ROUGE- & ROUGE- & ROUGE-& ACC = & BLEU =  \\ 
     \( + \mathrm{GEN_{rand}} \)  & EM =  61.7 & 1/2/L = & 1/2/L = & 1/2/L = & 78.2 & 10.1\\ 
     \( + \mathrm{REP_{gen}} \)  & & \!\!\! 19.0/3.6/13.1 \!\!\! &\!\!\! 23.0/6.1/15.6 \!\!\!&\!\!\! 34.7/12.0/27.1 \!\!\!& &  \\ \hline 
     \( \mathrm{DET_{NER}} \)  & F1 = 67.9 & ROUGE- & ROUGE- & ROUGE- & ACC = & BLEU =  \\ 
     \( + \mathrm{GEN_{prompt}} \)  & EM = 62.8 & 1/2/L = & 1/2/L = & 1/2/L = & 81.6 & \textbf{10.5} \\ 
     \( + \mathrm{REP_{gen}} \)  & & \!\!\! 19.6/3.8/13.6 \!\!\! &\!\!\! 24.1/6.6/16.1 \!\!\! &\!\!\! 34.3/11.6/26.7 \!\!\! & &  \\ \hline
     \( \mathrm{DET_{tag\_mask}} \)  & F1 = 74.1 & ROUGE- & ROUGE- & ROUGE- & ACC = & BLEU =  \\ 
     \( + \mathrm{GEN_{prompt}} \)  & EM = 70.6 & 1/2/L = & 1/2/L = & 1/2/L = & 80.8 & 6.9 \\ 
     \( + \mathrm{REP_{gen}} \)  & & \!\!\! 21.9/4.7/15.2 \!\!\! & \!\!\! \textbf{29.7}/\textbf{9.7}/\textbf{20.1} \!\!\! & \!\!\! 40.8/15.0/31.7 \!\!\!& &  \\ \hline 
     \( \mathrm{DET_{tag\_rep}} \)  & F1 = 71.3 & ROUGE- & ROUGE- & ROUGE- & ACC = & BLEU =  \\ 
     \( + \mathrm{GEN_{prompt}} \)  & EM = 66.8 & 1/2/L = & 1/2/L = & 1/2/L = & 81.6 & 8.0 \\ 
     \( + \mathrm{REP_{gen}} \)  & & \!\!\! 20.5/3.8/14.0 \!\!\! & \!\!\! 19.8/5.0/13.8 \!\!\! & \!\!\! 40.4/14.9/31.5 \!\!\! & &  \\ \hline 
    \end{tabular}
    \caption{Performance of various pseudonymization methods across different NLP tasks and datasets. The bolded parts in the table represent \textbf{the best results excluding the large-scale LLM}.}
    \vspace{-0.2cm}
    \label{tab:exp_result}
\end{table*}

\paragraph{Datasets}
\label{sec:dataset}

We conduct experiments on several publicly available real-world datasets across various NLP tasks, including SQuAD 2.0~\citep{rajpurkar-etal-2016-squad} for question answering, XSum~\citep{Narayan2018DontGM}, CNN/Dailymail~\citep{see-etal-2017-get}, and SAMSum~\citep{gliwa-etal-2019-samsum} for summarization, GLUE (MNLI)~\citep{williams2017broad, wang2019glue} for natural language inference, and WMT14 (de-en)~\citep{bojar-EtAl:2014:W14-33} for machine translation. 
For experimental efficiency, we randomly sampled 1,000 samples from the test sets of each dataset to serve as the test set. 
In this study, we focus our analysis on three primary categories of named entities: person, location, and organization. 

\paragraph{Evaluation Metrics}
\label{sec:metrics}
For different datasets, we will use distinct performance evaluation metrics. 
For SQuAD 2.0, we use the F1 score and Exact Match (EM)~\citep{Rajpurkar2018SQuAD2} as the evaluation metrics. 
For XSum, CNN/Dailymail, and SAMSum, we use ROUGE-1, ROUGE-2, and ROUGE-L~\citep{lin2004rouge} as the evaluation metrics. 
For GLUE (MNLI), we use the accuracy score as the evaluation metric. 
For WMT14 (de-en), we use the BLEU-4~\citep{papineni2002bleu} score as the evaluation metric. 
In addition to these performance evaluation metrics, we also calculate the distance between the original text \( X \) and the replaced text \( X' \), defined as \( 1 - s(X, X') \), to assess the effectiveness of the pseudonymization method. 
Here, \( s(X, X') \) represents the cosine similarity between the sentence embedding vectors of \( X \) and \( X' \), both of which are computed using a pretrained model, All-Mpnet-Base-V2~\footnote{
\url{https://huggingface.co/sentence-transformers/all-mpnet-base-v2}
}.

\paragraph{Baseline Methods}
\label{sec:baseline}
We designed two baseline methods and compared the pseudonymization method described in this paper with these baselines: (1) directly using a cloud-based LLM (simulated using a locally deployed large-scale LLM) to perform experimental NLP tasks, and (2) directly using a local small-scale instruction-tuned LLM to perform experimental NLP tasks.

\begin{figure*}[t]
    \centering
    \includegraphics[scale=0.36]{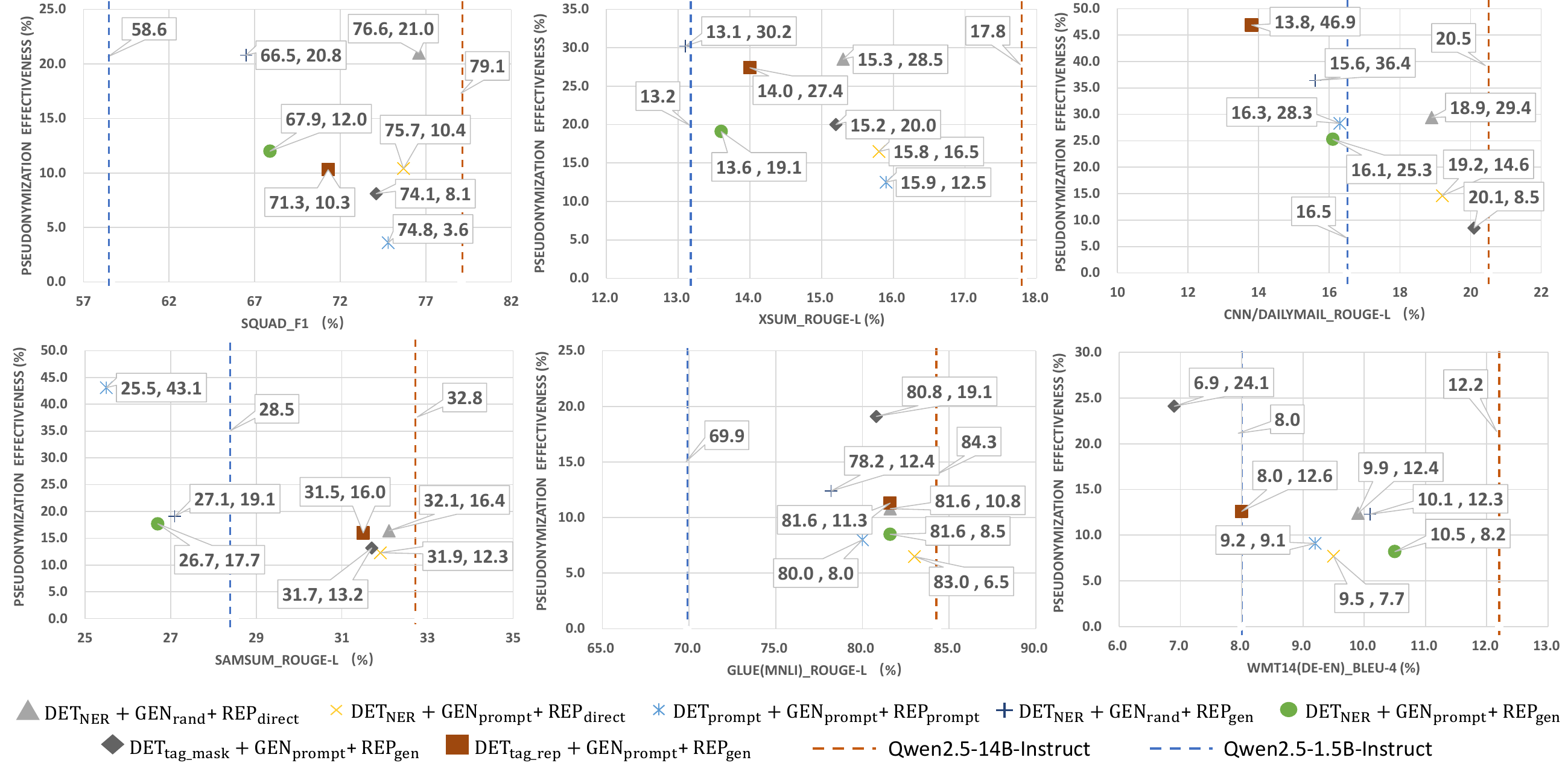}
    \caption{Performance metrics and pseudonymization effectiveness of various methods across different datasets}
    \vspace{-0.2cm}
    \label{fig:plots}
\end{figure*}

\paragraph{Implementation Details}
\label{sec:impl}
For the efficiency of the experiments, we locally deployed the Qwen2.5-14B-Instruct\footnote{
\url{https://huggingface.co/Qwen/Qwen2.5-14B-Instruct}
} as the large-scale LLM to simulate the cloud-based LLMs. 
We used the Qwen2.5-1.5B-Instruct\footnote{
\url{https://huggingface.co/Qwen/Qwen2.5-1.5B-Instruct}
} as the local small-scale instruction-tuned LLM for the prompt-based detection, generation, and replacement methods.
As described in Section~\ref{sec:pesu_method}, we then fine-tuned the Qwen2.5-1.5B model\footnote{ \url{https://huggingface.co/Qwen/Qwen2.5-1.5B} } to output either a repetition of the input text or the results of the Seq2Seq detection method for executing the replacement approach through controllable text generation. 
A total of 20,000 samples were randomly selected from the training sets of each dataset. 
Following the procedure outlined in Table~\ref{tab:exp_seq2seq_det}, these samples were preprocessed and subsequently used as fine-tuning data. 
We fine-tuned the Qwen2.5-1.5B model for 3 epochs using a learning rate of 1.0e-4.
%

\subsection{Main Result}
\label{sec:result}

Notably, each component of the proposed pseudonymization framework is decoupled, allowing the methods described in Section~\ref{sec:framework} to be freely combined. 
We evaluate the majority of possible method combinations and present the results of several representative approaches, comparing them against the baselines. 
The results are shown in Table~\ref{tab:exp_result}.
It is evident that across various NLP tasks and datasets, pseudonymization methods based on the proposed framework achieve results comparable to those of the large-scale LLM baseline. 
Specifically, these methods achieve over 95\% of the large-scale LLM baseline's performance on SQuAD 2.0, CNN/DailyMail, SAMSum, and GLUE (MNLI), over 90\% on XSum, and approximately 85\% on WMT14 (de-en).
Across all datasets, the proposed methods significantly outperform the small-scale LLM baseline.
It is important to note that, in real-world scenarios, the parameter scale of cloud-based models is expected to be significantly larger than that of the locally deployed large-scale LLM baseline. 
This further highlights the necessity of the pseudonymization framework proposed in this paper for enabling the secure remote use of cloud-based large-scale LLMs.

\begin{table*}[th]
    \centering
    \begin{tabular}{|l||p{12cm}|}
    \hline
         \textbf{Premise} & The vineyards hug the gentle slopes between the \textbf{Vosges} and the \textbf{Rhine Valley} along a single narrow 120-km (75-mile) strip that stretches from \textbf{Marlenheim}, just west of \textbf{Strasbourg}, down to \textbf{Thann}, outside \textbf{Mulhouse}. \\ \hline
         \textbf{Hypothesis} & The slopes between the \textbf{Vosges} and \textbf{Rhine Valley} are the only place appropriate for vineyards. \\ \hline
         \textbf{Answer} & neutral  \\ \hline \hline
         \textbf{Large-scale LLM} & neutral (correct) \\ \hline \hline
         \textbf{small-scale LLM} & contradiction (incorrect) \\ \hline \hline
         \!\!\!\begin{tabular}{l}
         \( \mathbf{DET_{NER}} \) \\
         \( + \mathbf{GEN_{rand}} \) \\
         \( + \mathbf{REP_{gen}} \) \\
         \end{tabular}\!\!\!  & \!\!\!\begin{tabular}{l|p{9.7cm}}
            Premise: & The vineyards hug the gentle slopes between the \textbf{Eifel Mountains} and the \textbf{Danube River Basin} along a single narrow 120-km (75-mile) strip that stretches from \textbf{Marsden}, just west of \textbf{Erlangen}, down to \textbf{Thompson}, outside \textbf{Lyon City}.  \\ \hline
            Hypothesis: & The slopes between the \textbf{Eifel Mountains} and \textbf{Danube River Basin} are the only place appropriate for vineyards. \\ \hline
            Answer: & neutral (correct) \\ 
        \end{tabular}\!\!\! \\ \hline         
    \end{tabular}
    \caption{Example of correct output by the proposed method on GLUE (MNLI) dataset compared to baselines}
    \label{tab:cor_exp}
\end{table*}

\begin{table}[th]
    \centering    
    \begin{tabular}{c}
        \begin{tabular}{l||cccc}
            \hline
            &  NER  &  prompt  & tag\_mask & tag\_rep \\ \hline \hline
           \!\!\!PRR\!\!\!& \textbf{65.7} & 47.9 & 33.5 & 43.1 \\ \hline
        \end{tabular}
        \\ (a) \\  \\
        \begin{tabular}{l||cc}
           \hline
           & rand & prompt \\ \hline \hline
           PPS & \textbf{74.9} & 45.2 \\ \hline
        \end{tabular}
    \\ (b) \\ \\
        \begin{tabular}{l||ccc}
        \hline
           & direct & prompt & gen \\ \hline \hline
           SCS & 20.9 & 19.7 & \textbf{19.2} \\ \hline
        \end{tabular}
    \\ (c) \\ 
    \end{tabular}
    \caption{(a) Privacy Removal Rate (PRR) for each detection method. (b) Privacy Preservation Score (PRS) for each generation method. (c) Semantic Correctness Score (SCS) for replacement method.}
    \label{tab:anay_result}
\end{table}

We further compared the key performance metrics and pseudonymization effectiveness of each method across different NLP tasks and datasets, with the results visualized in Figure~\ref{fig:plots}.
An interesting finding is that, in tasks like QA and summarization, which are less reliant on the semantic details of the text, the combination of \( \mathrm{DET_{NER}} + \mathrm{GEN_{rand}} + \mathrm{REP_{direct}}\) achieves the best overall results in both performance metrics and pseudonymization effectiveness. 
However, in tasks like MNLI and MT, where text details significantly impact the results, the combination of \( \mathrm{DET_{NER}} + \mathrm{GEN_{rand}} + \mathrm{REP_{gen}}\) and \( \mathrm{DET_{tag_mask}} + \mathrm{GEN_{prompt}} + \mathrm{REP_{gen}}\) consistently yields the best overall performance.

Table~\ref{tab:cor_exp} presents an example of the correct output generated by the proposed method.
In this example, entities in the premise and hypothesis texts, such as ``Vosges'' and ``Rhine Valley'', were replaced with other entities, like ``Eifel Mountain'' and ``Danube River Basin'', using the combination of \( \mathrm{DET_{NER}} + \mathrm{GEN_{rand}} + \mathrm{REP_{gen}}\). 
This effectively protects the potential privacy information contained within those entities. 
Meanwhile, when the pseudonymized text was processed by a large-scale LLM, it generated the correct inference, whereas the small-scale model failed to do so.

\subsection{Discussion}
\label{sec:anal}
We further evaluated the effectiveness of various methods in achieving the stage-specific objectives throughout the different stages of the proposed pseudonymization framework.

First, we calculate the Privacy Removal Rate (PRR) for each privacy entity detection method using the formula \( \mathrm{PRR} = \frac{\mathrm{card}(P' \cap P)}{\mathrm{card}(P)} \times 100(\%) \), where \( \mathrm{card}(\cdot) \) denotes the cardinality of the corresponding set. 
The results are shown in Table~\ref{tab:anay_result} (a). 
Notably, the NER-based detection method yielded the highest PRR.

We compute the Privacy Preservation Score (PPS) for each replacement candidate generation method as the average distance between \( p'^j_{A_{i}} \) and \( q^j_{A_{i}} \), following the formula \( \mathrm{PPS} = \mathrm{avg}(1 - s(p'^j_{A_{i}}, q^j_{A_{i}})) \times 100 (\%) \). 
It is evident that a higher PPS score indicates greater difficulty in inferring the privacy entity from the replacement entity, thereby offering better protection for privacy information. 
The results are presented in Table~\ref{tab:anay_result} (b). 
Notably, the random sampling generation method achieved the highest PPS.

We compute the Semantic Correctness Score (SCS) to assess the effectiveness of each entity replacement method by measuring the perplexity of \(X'\) using Qwen2.5-1.4B-Instruct. 
The SCS is calculated as \( \mathrm{SCS} = \mathrm{avg}(\mathrm{loss}(f(x'_{<i}),x'_{i})) \) (\(x'_{i} \in X'\)), where \( f(\cdot) \) represents the next-token prediction function, and \( \mathrm{loss}(\cdot) \) denotes the loss function of the language model. 
A lower SCS indicates that \(X'\) better aligns with the probability distribution of the language model, thereby exhibiting higher semantic correctness. 
The results are presented in Table~\ref{tab:anay_result} (c). 
Notably, replacement through controllable text generation achieved the lowest SCS.

\section{Conclusion}
\label{sec:con}
In this paper, we outline the process of remote user interaction with LLMs and propose a comprehensive definition of a pseudonymization framework applicable to cloud-based LLMs. 
We believe that this framework provides a universally applicable approach to the text pseudonymization process and can serve as a guide for future research in this area.
Additionally, we introduce a pseudonymization method based on a controllable text generation process, which ensures that the replaced text maintains maximal semantic correctness. 
Experimental results demonstrate that the proposed framework strikes an optimal balance between privacy protection and utility.

\section*{Limitations}

The primary limitation of this work is that the pseudonymization process is implemented through three relatively independent processing stages rather than an end-to-end machine learning approach. However, even end-to-end pseudonymization methods must inherently incorporate the three stages outlined in this paper: detection, generation, and replacement. Given that these stages have distinct problem definitions and task objectives, integrating them into a unified end-to-end framework presents a significant challenge. Addressing this challenge will be a key focus of our future research.

In addition, we utilized straightforward methods to accomplish the objectives of each stage, such as NER and prompt-based approaches. However, the primary contribution of this work lies in proposing a general pseudonymization framework. 
Within this framework, incorporating more advanced methods at each stage is expected to enhance overall performance.

For the sake of experimental efficiency, this work employs the same entity replacement method in both the restoration and pseudonymization processes. However, in practical applications, different replacement methods could be utilized for these two processes, potentially enhancing the overall effectiveness of the approach.

Although this work has validated the effectiveness of the proposed framework and methods on multiple NLP tasks across different datasets, certain tasks, such as text continuation, remain unexplored. Text continuation presents unique challenges for pseudonymization and restoration, as it may generate entities not present in the input text. Future work will include experiments to address this aspect.

\bibliography{custom}

\end{document}